\documentclass[aps,prb,twocolumn,superscriptaddress,showpacs]{revtex4-1}

\usepackage{graphicx}
\usepackage{calc}
\usepackage{bm}
\usepackage{color}
\usepackage{mathtools}

\begin{document}

\title{Magnetocaloric effect for the altermagnetic candidate  MnTe}

\author{N.N. Orlova}
\author{V.D. Esin}
\author{A.V.~Timonina}
\author{N.N.~Kolesnikov}
\affiliation{Institute of Solid State Physics of the Russian Academy of Sciences, Chernogolovka, Moscow District, 2 Academician Ossipyan str., 142432 Russia}
\author{E.V.~Deviatov}
\affiliation{Institute of Solid State Physics of the Russian Academy of Sciences, Chernogolovka, Moscow District, 2 Academician Ossipyan str., 142432 Russia}
\affiliation{V.L. Ginzburg Research Centre for High-Temperature Superconductivity and Quantum Materials, P.N. Lebedev Physical Institute of RAS, Moscow 119991, Russia}

\date{\today}

\begin{abstract}
We experimentally investigate magnetocaloric effect for single crystals of altermagnetic candidate MnTe at the transition to the  state with spontaneous spin polarization, i.e. well below the N\'eel temperature of MnTe. The isothermal magnetic  entropy change $\Delta S$ is calculated from the experimental magnetization curves by using Maxwell relation. We observe well-defined magnetocaloric effect as a narrow $\Delta S$ peak around the critical temperature $T_c\approx 81$~K, which  reflects the sharp magnetization jump.  This behavior is unusual for standard ferromagnetic transitions, so it  confirms  the predicted spin-orbit-induced spin polarization in the MnTe altermagnetic state and, therefore, confirms MnTe as a prototypical altermagnet candidate. 
\end{abstract}

\maketitle

\section{Introduction}

Recently, a new class of altermagnetic materials has been proposed~\cite{alter_common,alter_mazin}.  The   usual  ferro- and antiferro- magnetic classes are not characterized by  noticeable spin-orbit coupling, i.e. they belong to non-relativistic groups of magnetic symmetry. In contrast, one of the central concepts of the topological materials is the spin-momentum locking~\cite{sm-valley-locking}, which, usually, appears due to spin-orbit coupling~\cite{Armitage}. In altermagnets, the concept of spin-momentum locking was extended to the case of weak spin-orbit coupling, e.g. in collinear antiferromagnets. As a result, altermagnets are characterized by alternating spin splitting in the k-space~\cite{alter_common,alter_josephson}, so the up-spin-polarized subband and the down-polarized one are connected by ($\pi/2$, $\pi/3$, etc) rotation in the reciprocal space~\cite{alter_supercond_notes,alter_normal_junction}.

Despite the initial theoretical proposal,  spin-orbit coupling still can be important in real altermagnetic materials. For example, observation of the anomalous Hall effect (AHE)~\cite{Armitage} is regarded~\cite{alter_original,AHE_k_topol} as one of the main experimental demonstrations of  altermagnetism for non-ferromagnetic materials~\cite{AHE_MnTe1,AHE_MnTe2,AHE_Mn5Si3}. However, it is a common  agreement, that AHE still requires finite  net magnetization~\cite{satoru}, which appears due to the finite spin-orbit coupling  in altermagnets~\cite{spin_ferro_soc}. 

\begin{figure}
\includegraphics[width=1\columnwidth]{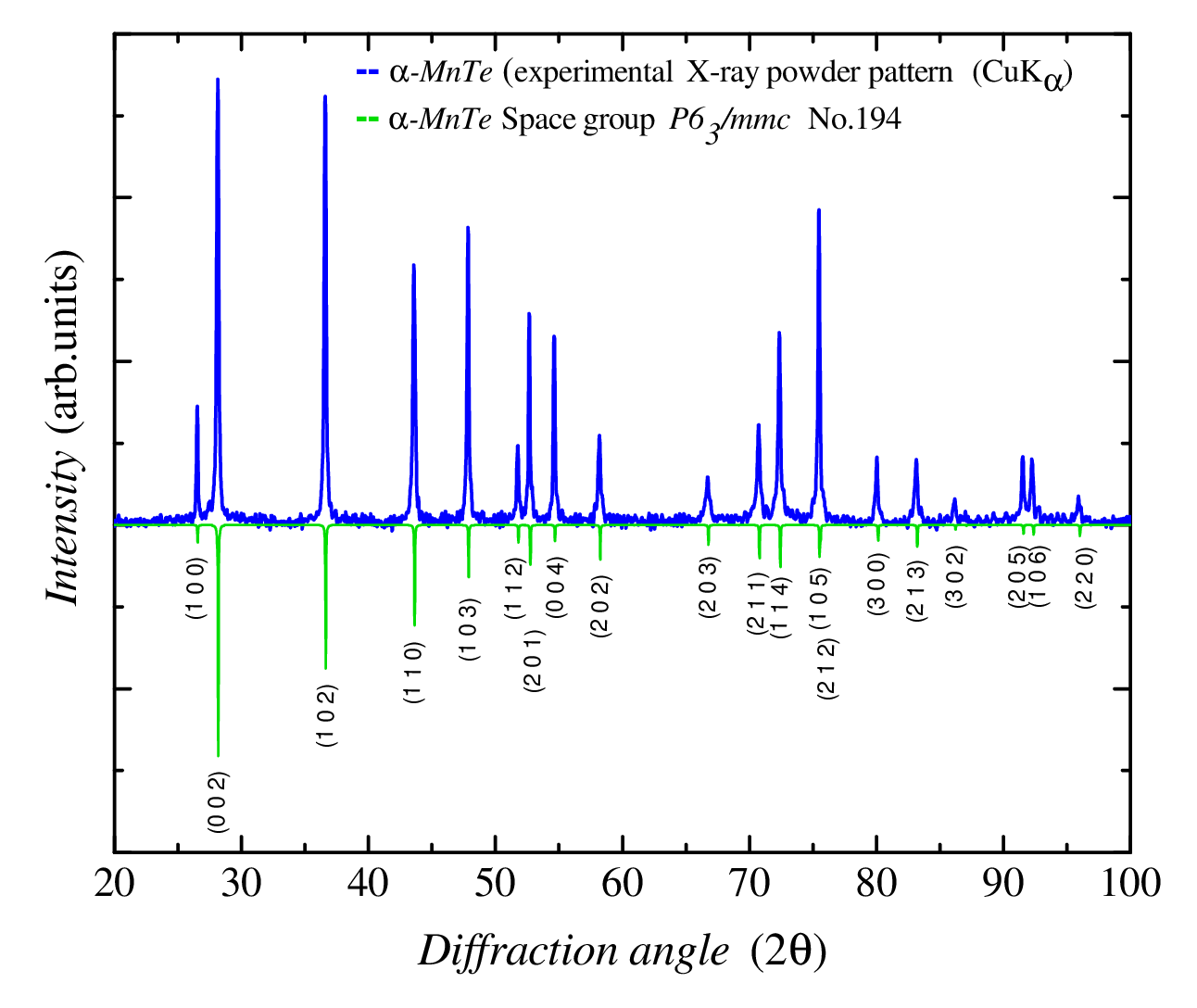}
\caption{(Color online) The X-ray powder diffraction  pattern (Cu K$_{\alpha}$ radiation), which is obtained for the crushed MnTe single crystal. The single-phase  $\alpha$-MnTe is confirmed with the space group $P6_3 /mmc$ No. 194, so the results below cannot appear from the incorrect stoichiometry or oxides. 
  }
\label{fig0}
\end{figure}

MnTe is regarded as a prototypical altermagnet candidate~\cite{MnTe_Mazin}, while for years it was known as a collinear antiferromagnet~\cite{MnTe1,MnTe2,MnTe3,MnTe4,MnTe5,MnTe6}. The magnetic moments on Mn have a parallel alignment within the $c$ planes and an antiparallel alignment between the planes, so the two magnetic sublattices are connected by a sixfold screw axis along [0001] in real space. At low temperatures, MnTe shows spontaneous AHE signals at zero magnetic field~\cite{AHE_MnTe1,AHE_MnTe2} and spontaneous ferromagnetic-like magnetization~\cite{orlova_mnte,orlova_mnte1}, which are the main experimental confirmations of altermagnetism in MnTe. Since spontaneous  magnetization still requires spin-orbit coupling, it has been  investigated in MnTe by temperature-dependent angle-resolved photo-emission spectroscopy (ARPES) and by disordered local moment calculations~\cite{MnTe_SO}. It seems to be important, that spin-orbit coupling is well-resolved~\cite{MnTe_SO} only below 100~K,  which is consistent with appearance of spontaneous magnetization~\cite{orlova_mnte,orlova_mnte1}. 
 At this transition, the easy magnetization axis is  $\pi/2$ rotated~\cite{orlova_mnte1} either by increasing the field above 1~kOe or the temperature above 81~K. The observed behavior cannot be expected for antiferromagnetics, thus, it requires to take into account the formation of the altermagnetic ground state for MnTe single crystals.

Generally speaking, altermagnetism presents a new type of spin ordering, which should be detectable by observation of the magnetocaloric effect (MCE). The latter is known  as the entropy change $\Delta S$ while the system undergoes the transition between differently spin ordered states~\cite{CoSnS_coloric}.  MCE is  of special interest due to the possible applications~\cite{MCEmater1,MCEmater2}, for example,  utilizing of antiferromagnetic coupling should increase energy efficiency of a MCE refrigerator. For the fundamental research, irrespective of the entropy change values, observation of MCE should confirm transition between differently spin-orderd states in altermagnetic materials. For example, it is reasonable to study MCE at the transition to the spin-orbit-induced~\cite{MnTe_SO} spin-polarized  state~\cite{AHE_MnTe1,AHE_MnTe2,orlova_mnte,orlova_mnte1} around 81~K in altermagnetic candidate  MnTe.

Here, we experimentally investigate magnetocaloric effect for single crystals of MnTe altermagnet at the transition to the  state with spontaneous spin polarization, i.e. well below the N\'eel temperature of MnTe. We observe well-defined magnetocaloric effect as a narrow $\Delta S$ peak around the cricital temperature $T_c\approx 81$~K, which reflects the sharp magnetization jump.

\begin{figure}
\includegraphics[width=1\columnwidth]{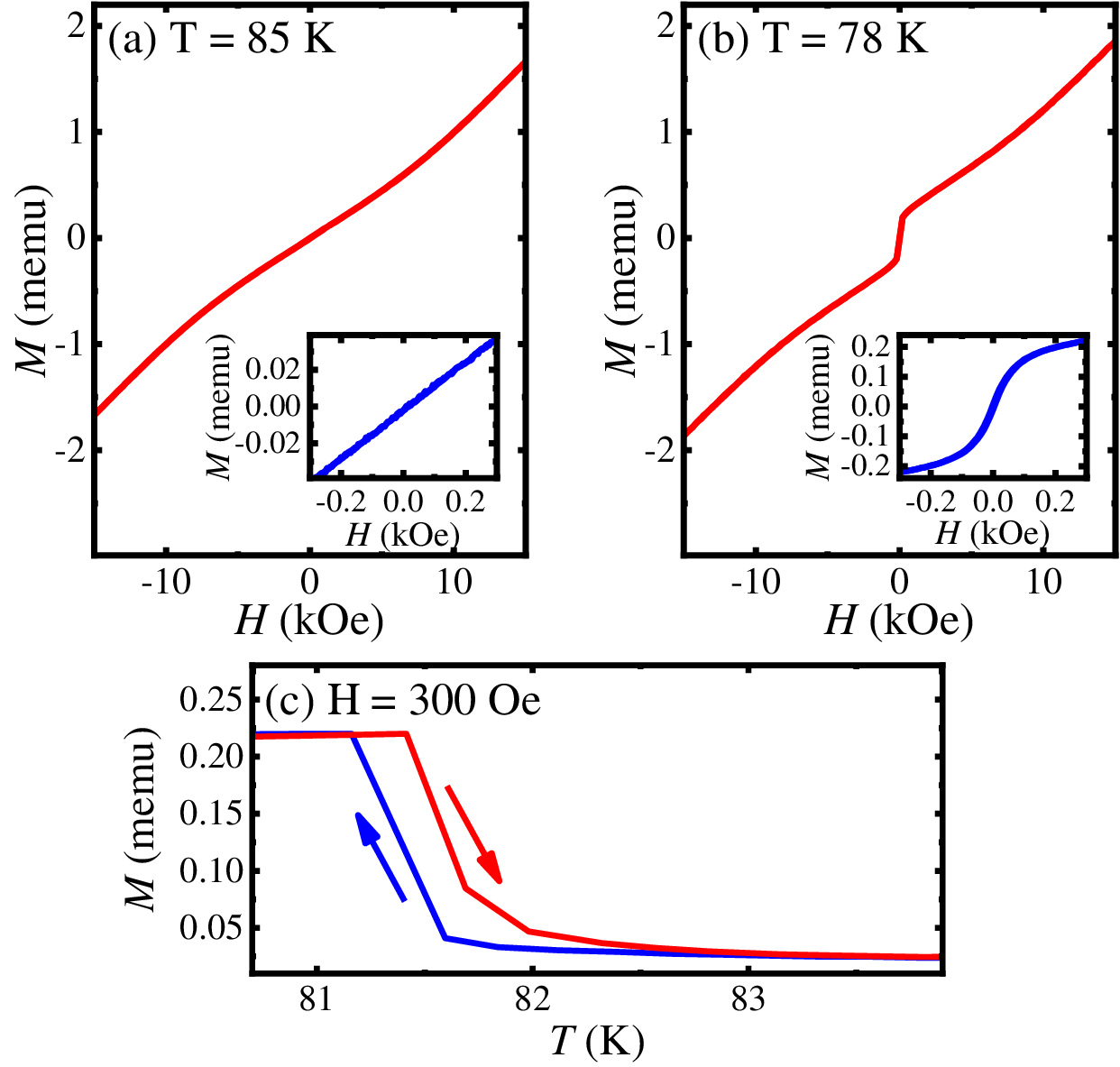}
\caption{(Color online) Transition from antiferromagnetic to the ferromagnetic-like magnetization around 81~K for the 5.73~mg MnTe sample, see Refs.~\cite{orlova_mnte,orlova_mnte1} for details.  
(a) $M(H)$ magnetization curves above the transition, at 85~K temperature. There are nonlinear $M(H)$ branches in high magnetic fields,  while $M(H)$ is strictly linear in low fields, as it is depicted in the inset. This $M(H)$ behavior is expected for the antiferromaghetic state
(b) Nonlinear $M(H)$ branches are shifted vertically at low 78~K temperature, so a pronounced zero-field kink can be seen. The inset shows low-field $M(H)$ ferromagnetic-like  hysteresis. 
(c) Appearence of spontaneous magnetization below 81~K as a sharp $M(T)$ jump. The red and the blue curves are obtained at 300~Oe magnetic field, for heating and cooling, respectively.
  }
\label{fig1}
\end{figure}

\section{Samples and technique}

$\alpha$-MnTe was synthesized by reaction of elements (99.99\% Mn  and 99.9999\% Te) in evacuated silica ampules slowly heated up to 1050--1070$^\circ$C. The obtained loads were melted in the graphite crucibles under 10 MPa argon pressure, then homogenized at 1200$^\circ$C for 1 hour. The crystals grown by gradient freezing method are groups of single crystal domains with volume up to 0.5--1.0~cm$^3$. The MnTe composition is verified by energy-dispersive X-ray spectroscopy.  The powder X-ray diffraction analysis confirms single-phase $\alpha$-MnTe with the space group $P6_3 /mmc$ No. 194, see Fig.~\ref{fig0}. As usual, the powder has been obtained by crashing the MnTe single crystal. High quality of our MnTe single crystals is also confirmed by high reproducibility  of the results, which are obtained for MnTe flakes of different mass and shape~\cite{orlova_mnte,orlova_mnte1}.

The  measurements are performed by Lake Shore Cryotronics 8604 VSM magnetometer, equipped with nitrogen flow cryostat. We investigate two mechanically cleaved single crystal MnTe flakes of different mass (2.89~mg and 5.73~mg, respectively) to confirm reproducibility  of the results. A flake is mounted to the magnetometer sample holder by a low temperature grease, which has been tested to have a negligible magnetic response~\cite{GeTe_magnetization}. Small single crystal flakes are preferable for sensible measurements, while, e.g. for thin films the results may be seriously affected by the geometrical factors and by the admixture of the substrate magnetic response, especially in low magnetic fields~\cite{MnTe_film_6phi}. The sample can be rotated in magnetic field, so the hard and easy magnetization directions (i.e. sample orientation)  can be obtained~\cite{orlova_mnte,orlova_mnte1} from angle-dependent magnetization $M(\alpha)$.

We investigate the magnetocaloric effect as the isothermal magnetic  entropy change $\Delta S$ which is calculated from the experimental magnetization curves by using Maxwell relation~\cite{DeltaS,technique,CoSnS_coloric}:
 
\begin{equation}
\Delta S = \mu_0\int^H_0 \left(\frac{\partial M}{\partial T}\right)_H dH \label{DeltaS}
\end{equation}

As a first step, a set of $M(H,T=const)$ isotherms is obtained for the required magnetic field range. Subsequently, the first derivative $\left(\frac{\partial M}{\partial T}\right)_H$ is calculated as a result of subtraction $\left(\frac{\Delta M}{\Delta T}\right)(H,T)$ of every two neighbor curves. As a final step, $\Delta S$ is achieved for every temperature point by  integration over the magnetic field range~\cite{technique,CoSnS_coloric}.

\section{Experimental results}

\begin{figure}
\includegraphics[width=\columnwidth]{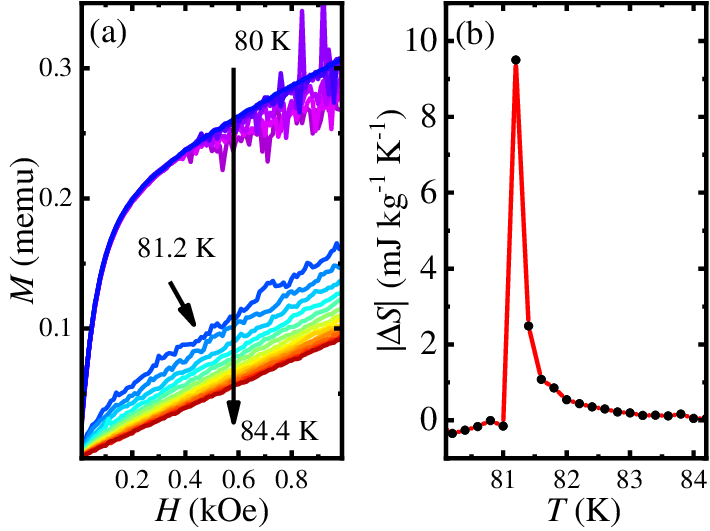}
\caption{(Color online)  (a) $M(H,T=const)$ isothermal magnetization curves from 80~K to 84.4~K temperatures with 0.2~K step, with sharp $M(T)$ jump at the transition temperature $T_c\approx 81$~K. The initial sample state is obtained after ZFC from room temperature to 78~K. Between the curves, temperature is always stabilized at zero field. 
(b) The isothermal magnetic entropy change $\Delta S$ with narrow peak at the critical temperature $T_c\approx 81$~K. The obtained  $|\Delta S|$ values are much below the usual ones for standard ferromagnetic transitions around the Curie temperature~\cite{CoSnS_coloric}. The data are obtained for the 5.73~mg  MnTe sample.
 }
\label{fig2}
\end{figure} 

Fig.~\ref{fig1} shows appearance~\cite{orlova_mnte,orlova_mnte1} of the spontaneous ferromagnetic-like magnetization around 81~K for the 5.73~mg MnTe sample. $M(H)$ magnetization dependences are measured for easy magnetization direction~\cite{orlova_mnte,orlova_mnte1} by standard method of the magnetic field gradual sweeping between two opposite field values. 

Let us start from higher temperatures, i.e. from the well-known antiferromagnetic state. 
At 85~K temperature, there are nonlinear $M(H)$ branches in high magnetic fields, see Fig.~\ref{fig1} (a), while $M(H)$ is strictly linear in low fields, as it is demonstrated in the inset. The linear $M(H)$ is expected for the antiferromaghetic state, while the high-field nonlinear $M(H)$ branches are due to the antiferromagnetic domain configuration change  below the N\'eel vector reorientation field~\cite{MnTe_film_6phi,AFM_book}. 

At low 78~K temperature, there are also nonlinear $M(H)$ branches, but they are shifted vertically, so a pronounced zero-field kink can be seen in Fig.~\ref{fig1} (b). The low-field $M(H)$ curves are shown in the  inset to Fig.~\ref{fig1} (b), they are obtained with high accuracy for smaller magnetic field step for two sweep directions. The inset confirms ferromagnetic-like  hysteresis, i.e. spontaneous magnetization, at this temperature~\cite{orlova_mnte,orlova_mnte1}.

The detailed $M(T)$ temperature dependence is shown in Fig.~\ref{fig1} (c). The temperature is stabilized with  0.2~K step from 78.4~K to 83.8~K. The red and the blue curves are obtained at 300~Oe magnetic field, for heating and cooling, respectively.   It can be seen from the curves, that ferromagnetic-like spontaneous magnetization appears below 80.5--81.5~K as a sharp $M(T)$ jump.  

As the main experimental result, we investigate magnetocaloric effect around the transition to the ferromagnetic-like state. 
$M(H,T=const)$ isothermal magnetization curves are obtained in a temperature range, that  covers the transition temperature $T_c\approx 81$~K, see Fig.~\ref{fig2} (a). Before any set of  $M(H,T=const)$ isotherms, it is necessary to set up the stable initial magnetic state of the sample~\cite{CoSnS_coloric}. In our experiment, it is performed by cooling the sample in zero magnetic field (ZFC protocol) from the room temperature to 78~K one, and subsequent sample magnetization along the easy axis by sweeping the field between -15~kOe and +15~kOe values. Afterward, temperature is stabilized with 0.2~K step, $M(H)$ curves are obtained from 80~K to 84.4~K, see Fig.~\ref{fig2} (a). Between the curves, temperature is always stabilized at zero field. 

The isothermal magnetic entropy change $\Delta S$ can be calculated by Eq.(\ref{DeltaS}) , absolute $|\Delta S|$ values are presented in Fig.~\ref{fig2} (b). As it should be expected from sharp $M(T)$ jump in Fig.~\ref{fig1} (c) and in Fig.~\ref{fig2} (a), the obtained $|\Delta S|$ shows narrow peak at the cricital temperature $T_c\approx 81$~K in Fig.~\ref{fig2} (b). MnTe heat capacity $C_p$ equals~\cite{CpMnTe1,CpMnTe2} to 191.3~J$\times$kg$^{-1}\times$K$^{-1}$ for 80~K, so one can estimate~\cite{DeltaTad,DeltaS} $\Delta T = -\frac{T\Delta S}{C_p} \approx 4$~mK at the transition point. Thus, the obtained  $|\Delta S|$ and $\Delta T$ values are much below the usual ones for standard ferromagnetic transitions around the Curie temperature~\cite{CoSnS_coloric}. 

 The $M(T)$ transition is unusually sharp in our case, which can be directly seen from   the original  $M(H)$ curves in Fig.~\ref{fig2} (a). The difference between the temperature points is limited by requirement of temperature stabilization,  the quality of stabilization is also confirmed by  the original  M(H) curves in Fig.~\ref{fig2} (a). However, there are still several experimental point for the peak region in Fig.~\ref{fig2} (b). Moreover, the peak's shape well reflects the  $M(H)$ curves evolution: the jump in $|\Delta S|$ is sharp between 81K and 81.2K, while the drop is smoother above 81.2K in Fig.~\ref{fig2} (b).

\begin{figure}[t]
\includegraphics[width=\columnwidth]{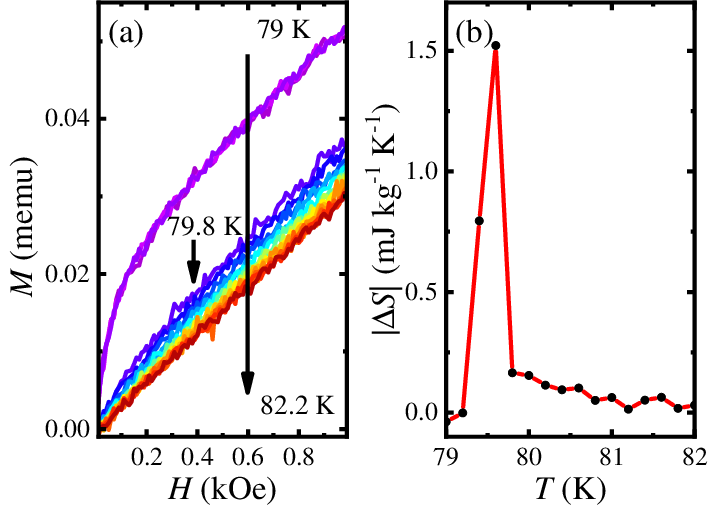}
\caption{(Color online) $M(H,T=const)$ isothermal magnetization curves (a) and $|\Delta S|$ values (b) for the 2.89~mg  MnTe sample. The transition can be seen as sharp $M(T)$ jump at $T_c\approx 79.5$~K, which well corresponds to the position of the $|\Delta S|$ peak. The peak  amplitude is diminished for this 2.89~mg  MnTe sample in comparison with the 5.73~mg one in Fig.~\ref{fig2} (b). The magnetization value does not scale with the sample mass for the ferromagnetic-like state in low fields~\cite{orlova_mnte,orlova_mnte1}.
 }
\label{fig3}
\end{figure}

The reported behavior is sample-independent, it can be reproduced for any MnTe single crystal sample. For example,  Fig.~\ref{fig3} shows $M(H,T=const)$ isothermal magnetization curves (a) and $|\Delta S|$ values (b) for the 2.89~mg  MnTe sample. For this sample, the temperature range is chosen as 79~K -- 82.2~K. The transition can be seen as sharp $M(T)$ jump at $T_c\approx 79.5$~K, which well corresponds to the position of the $|\Delta S|$ peak. The peak  amplitude is diminished to $|\Delta S(T)| = 1.5$~mJ$\times$kg$^{-1}\times$K$^{-1}$ for this 2.89~mg  MnTe sample, while it one order of magnitude greater for the 5.73~mg one in Fig.~\ref{fig2} (b). This difference is due to the fact, that  the magnetization $M(H)$ value does not scale with the sample mass for the ferromagnetic-like state in low fields in Figs.~\ref{fig2} (a) and~\ref{fig3} (a). In contrast, magnetization value is proportional to the sample mass in high fields and/or above the transition temperature $T_c$  (see Refs.~~\cite{orlova_mnte,orlova_mnte1} for these data).

\section{Discussion} \label{disc}

As a result,  we observe well-defined magnetocaloric effect  for MnTe altermagnet candidate as a sharp $|\Delta S|$ peak around 81~K, the effect is well reproducible for different samples with low variation of  the critical temperature $T_c$~K.

First of all, the critical temperature $T_c\approx 81$~K is well below the N\'eel temperature of MnTe, which exceeds the room value~\cite{MnTe1,MnTe2,MnTe3,MnTe4,MnTe5,MnTe6}. Thus, the alignment of  magnetic moments on Mn is stable in this temperature range and, therefore, can not be responsible for the observed $|\Delta S|$ peak around 81~K. 

Secondly, both sharp $M(T)$ jump and relatively low $|\Delta S|$ peak values are unusual for standard ferromagnetic-paramagnetic transitions, i.e.  around the Curie temperature, see  Ref.~\cite{CoSnS_coloric} as an example. Thus, the effect could not be connected with any imperfections (magnetic impurities), if they occurred in our MnTe samples. The sample quality is independently confirmed by energy-dispersive X-ray spectroscopy and powder X-ray diffraction analysis~\cite{orlova_mnte,orlova_mnte1}.

On the other hand, MnTe belongs to a new class of altermagnetic materials~\cite{MnTe_Mazin}. For  MnTe it is accepted, that the principle origin of nonzero net magnetic moment~\cite{AHE_MnTe1,AHE_MnTe2,orlova_mnte,orlova_mnte1}, and, therefore,  of weak remanence  magnetization~\cite{alter_ferro,spin_ferro_soc} is the spin-orbit coupling~\cite{satoru} in the valence orbitals~\cite{Dichroism}. It seems to be important, that spin-orbit coupling is well-resolved only below 100~K in Ref.~\cite{MnTe_SO}, which is consistent with appearance of narrow $|\Delta S|$ peak  in Figs.~\ref{fig2} and~\ref{fig3}.  

Thus, narrow $|\Delta S|$ peak at the critical temperature $T_c$  confirms  the transition in the spin ordering  from the spin-orbit-induced spin polarization in the MnTe altermagnetic state, and, therefore, confirms MnTe as a prototypical altermagnet candidate~\cite{MnTe_Mazin}.  

For possible applications of magnetocaloric effect in altermagnets, small amplitude of $\Delta S$ and, therefore, of  $\Delta T$, can only allow to consider MnTe as a model system, which should stimulate further material search.  However, observation of magnetocaloric effect in MnTe opens a new way to transfer from ferromagnetically ordered systems to the altermagnetic ones with higher reversibility and  smaller energy costs, which is confirmed by small width of the $|\Delta S|$ peak at the critical temperature $T_c$ in comparison with standard ferromagnetic-paramagnetic transitions.

\section{Conclusion}
As a conclusion, we experimentally investigate magnetocaloric effect for single crystals of MnTe altermagnet at the transition to the  state with spontaneous spin polarization, i.e. well below the N\'eel temperature of MnTe. The isothermal magnetic  entropy change $\Delta S$ is calculated from the experimental magnetization curves by using Maxwell relation. We observe well-defined magnetocaloric effect as a narrow $\Delta S$ peak around the cricital temperature $T_c\approx 81$~K, which is  reflects the sharp magnetization jump.  This behavior is unusual for standard ferromagnetic transitions, so it  confirms  the predicted spin-orbit-induced spin polarization in the MnTe altermagnetic state. Despite it is believed, that  utilizing of altermagnet should increase the energy efficiency of a  refrigerator, MnTe is characterized by small amplitude of the magnetocaloric effect. Thus,  MnTe can  can only be considered as a model system for magnetocaloric applications, which can stimulate further material search.

\section{Acknowledgments}

We wish to thank S.S~Khasanov for X-ray sample characterization and Vladimir Zyuzin for valuable discussions.  We gratefully acknowledge financial support  by the  Russian Science Foundation, project RSF-26-12-00001, https://rscf.ru/project/26-12-00001/

\end{document}